\documentclass[sigconf]{acmart}
\usepackage[ruled,linesnumbered]{algorithm2e}
\usepackage{enumitem}
\usepackage{booktabs}
\usepackage{xcolor}
\usepackage{caption}
\usepackage{subcaption}
\usepackage{amsmath}
\usepackage{threeparttable}
\usepackage{multirow}
\usepackage{pifont}
\newcommand{\cmark}{\ding{51}}%
\newcommand{\xmark}{\ding{55}}%
\usepackage{marvosym}
\usepackage[multiple]{footmisc}
\usepackage{bigfoot}

\DeclareNewFootnote{AAffil}[arabic]
\DeclareNewFootnote{ANote}[fnsymbol]
\newcommand{\nop}[1]{}
\AtBeginDocument{%
  \providecommand\BibTeX{{%
    \normalfont B\kern-0.5em{\scshape i\kern-0.25em b}\kern-0.8em\TeX}}}

\copyrightyear{2022} 
\acmYear{2022} 
\setcopyright{acmlicensed}\acmConference[WWW '22]{Proceedings of the ACM Web Conference 2022}{April 25--29, 2022}{Virtual Event, Lyon, France}
\acmBooktitle{Proceedings of the ACM Web Conference 2022 (WWW '22), April 25--29, 2022, Virtual Event, Lyon, France}
\acmPrice{15.00}
\acmDOI{10.1145/3485447.3511974}
\acmISBN{978-1-4503-9096-5/22/04}

\begin{document}

\title{RETE: Retrieval-Enhanced Temporal Event Forecasting on Unified Query Product Evolutionary Graph}



\author{Ruijie Wang$^{1\ast}$, Zheng Li$^{2\dagger}$, Danqing Zhang$^{2}$, Qingyu Yin$^{2}$, Tong Zhao$^{2}$, \\Bing Yin$^{2}$ and Tarek Abdelzaher$^{1}$}
\thanks{$^{\ast}$Part of work was done during internship at Amazon; $^{\dagger}$Corresponding author.}
\affiliation{%
  \institution{$^{1}$University of Illinois at Urbana Champaign, IL, USA $^{2}$Amazon.com Inc, CA, USA}
  \country{}
}

\email{{ruijiew2, zaher}@illinois.edu, {amzzhe, danqinz, qingyy, zhaoton, alexbyin}@amazon.com}

\renewcommand{\shortauthors}{Wang, et al.}

\begin{abstract}

With the increasing demands on e-commerce platforms, numerous user action history is emerging. Those enriched action records are vital to understand users' interests and intents. Recently, prior works for user behavior prediction mainly focus on the interactions with product-side information. However, the interactions with search queries, which usually act as a bridge between users and products, are still under investigated. In this paper, we explore a new problem named temporal event forecasting, a generalized user behavior prediction task in a unified query product evolutionary graph, to embrace both query and product recommendation in a temporal manner. To fulfill this setting, there involves two challenges: (1) the action data for most users is scarce; (2) user preferences are dynamically evolving and shifting over time. To tackle those issues, we propose a novel \textit{Retrieval-Enhanced Temporal Event} (RETE) forecasting framework. Unlike existing methods that enhance user representations via roughly absorbing information from connected entities in the whole graph, RETE efficiently and dynamically retrieves relevant entities centrally on each user as high-quality subgraphs, preventing the noise propagation from the densely evolutionary graph structures that incorporate abundant search queries. And meanwhile, RETE autoregressively accumulates retrieval-enhanced user representations from each time step, to capture evolutionary patterns for joint query and product prediction. Empirically, extensive experiments on both the public benchmark and four real-world industrial datasets demonstrate the effectiveness of the proposed RETE method.

\end{abstract}

%

\maketitle
\section{introduction}
\label{sec:intro}

On shopping websites or platforms, users type search queries and then perform various actions on the returned products. Such behaviors produce numerous interactions to both search queries and products, such as {\it ``type a search query''}, {\it ``click a product page''}, {\it ``add a product to the shopping cart''} or {\it ``purchase a product''}. Jointly modeling user-query and user-product interactions are essential to identifying users' preferences and intents for reasonable and interpretable behavior prediction. 

\begin{figure}[htb]
    \centering
    \includegraphics[width =0.95\linewidth]{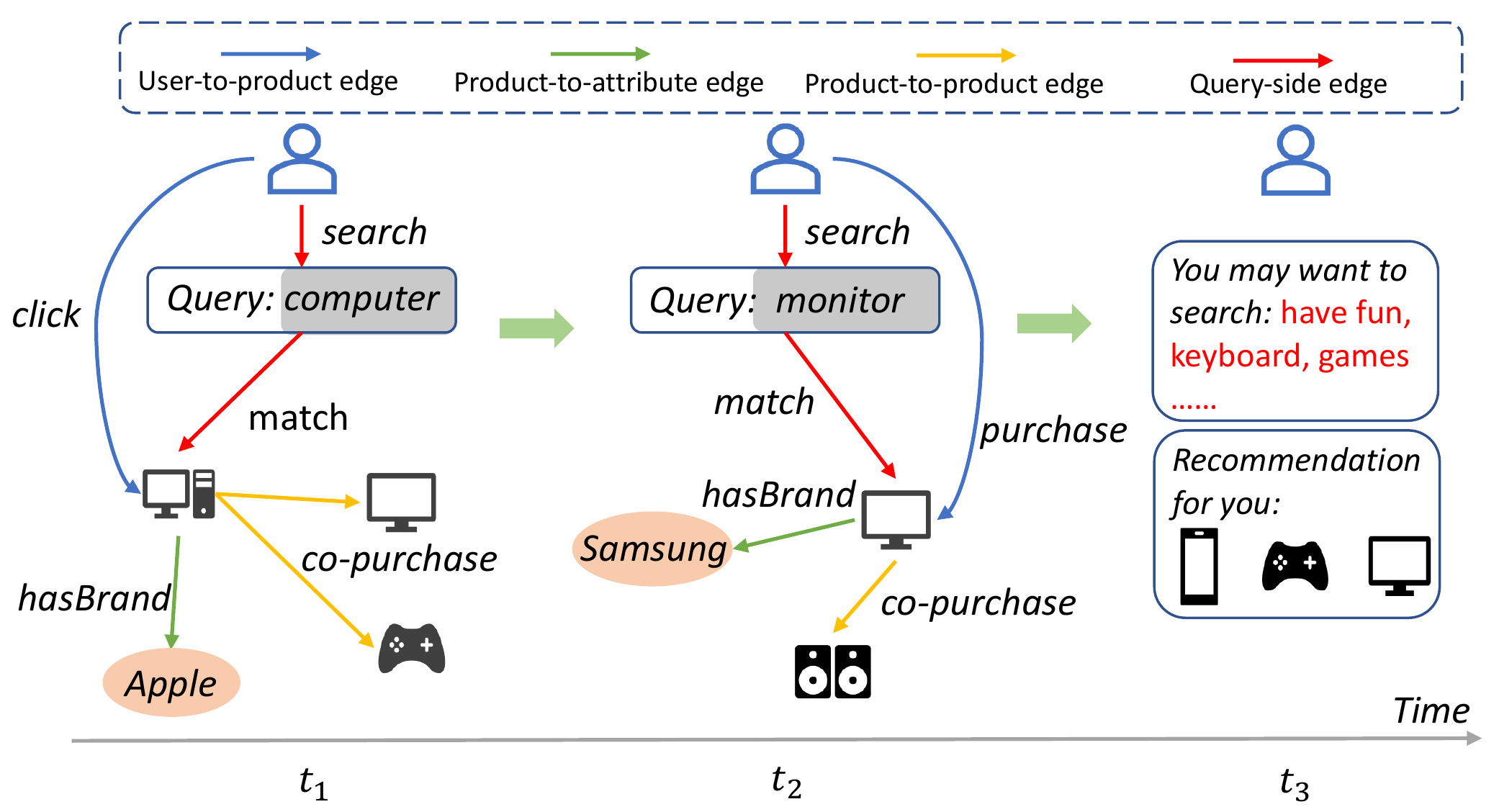}
    \caption{Illustration of temporal event forecasting on the E-commerce search domain. A set of interactions associated with temporal information between users, queries, products, and attributes constitute to user events at each time. Existing methods mainly focus on product-side interactions ({\color{blue}{Blue}}, {\color{yellow}{Yellow}}, {\color{green}{Green}} lines) while ignoring the underlying influence from search queries ({\color{red}{Red}} line) explored in our work.}
    \label{fig:task}
    \vspace{-2mm}
\end{figure}

Unfortunately, as shown in Figure~\ref{fig:task}, to predict user instant behaviors, existing works merely focus on product-side interactions (e.g, user-to-product~\cite{FM,NFM,CoEol,JODIE}, product-to-product~\cite{BERT4Rec,GRU4Rec,P-Companion}, or product-to-attribute~\cite{KGAT,ECFKG,CKE,RippleNet}) but neglect equally important search queries.  This usually causes incomplete and noisy user profile characterization, especially in the cold-start regime that lacks sufficient contexts. Concretely, as one crucial information carrier from human natural languages, search queries act like a bridge to link users and products. On one hand, users utilize search queries to express shopping intents, which are usually the prerequisite for subsequent user-product interactions. So queries can be directly closer to users' intents and suffice to reliably capture underlying interests. On the other hand, jointly modeling query and product information can bring in mutual benefits. For example, for users who accidentally interact with unrelated products in the imperfect matching sets returned by search engines, query information can be utilized to denoise those abnormal behaviors, and vice verses.

To embrace the collective power from both products and queries, we generalize user behavior prediction as a new problem setting named {\it Temporal Event Forecasting}, where a diverse range of interactions associated with temporal information between users, queries, products, and attributes constitute to a large-scale unified query product evolutionary graph. Our goal aims to dynamically capture user interests' dependency and evolution on this temporal knowledge graph by predicting user events at future times, where the event prediction can be disentangled as simultaneously predicting potential search queries and interacted products.

The challenges in fulfillment of the proposed setting are two-fold: (1) {\it scarce action data}: the limited action records of most users, due to the long-tail distribution~\cite{longtail}, make it hard to obtain robust user representations. To tackle the issue, prior works stack multiple aggregation layers~\cite{GAT,KGAT,RippleNet}, e.g., graph neural networks (GNNs)~\cite{gcn}, to accumulate information from other entities such as product side information~\cite{KGAT,ECFKG} or similar users~\cite{RippleNet}, to enhance user representations. However, as number of propagation layers grows, the neighborhood size increases exponentially, especially in a more connected graph like ours in light of abundant search queries. This can usually bring in extensive noises from unrelated and repeated entities, making user representations indiscriminative (a.k.a over-smoothing)~\cite{oversmoothing1,oversmoothing2,IMP}; (2) {\it interest evolution}: user preferences and intents are dynamically evolving and shifting over time. As new user activities are continuously emerging and collected, such new interaction events, reflecting users' most recent intents, may have large distribution gaps with previous user action behaviors. Failing to model such time-varing patterns usually results in  significant performance degradation as time goes by~\cite{retrain1,retrain2}, especially for modeling the joint objective of query and product.

Motivated by those challenges, we propose a retrieval-enhanced temporal event forecasting framework named {\it RETE}, to learn both discriminative and temporally-aware user representations for joint query and product recommendation. To enrich each user profile, RETE exploits subgraph samplers to dynamically filter out unrelated noises and retrieve higher-order entities centrally on each user from dense and temporal graph structures. Those retrieved entities are organized as subgraphs and integrated to enhance each user representation via a structural attention module. As such, the information propagation is preserved within the local related structures from high-quality subgraphs instead of the noisy global graph. To better capture user intent evolution over time, a sequence of retrieval-enhanced user representations from each time step are accumulated in an autoregressive manner via a temporal attention module, which can automatically learn the importance of user interaction events distributed in the axis of time. This adapts retrieval-enhanced user representation to be time-aware for capturing evolutionary patterns.

To validate the effectiveness of RETE, we conduct extensive experiments on both the public Yelp Challenge 2019 dataset and four real-world large-scale industrial datasets. Our experimental results demonstrate the superiority of RETE over state-of-the-art baselines by a large margin, including factorization machine (FM-based)~\cite{FM,NFM}, session-based~\cite{BERT4Rec,GRU4Rec}, knowledge graph (KG-based) ~\cite{KGAT,ECFKG} recommendation models as well as dynamic graph learning methods~\cite{JODIE}. We further conduct comprehensive ablation and hyper-parameter studies to validate the effectiveness of each design choices. Finally, we demonstrate the insights towards prediction interpretability by visualizing temporal weights across time steps.

Overall, our contributions can be summarized in follows:
\begin{itemize}[leftmargin = 15pt]
    \item \textbf{Problem formulation}:  To the best of our knowledge, our work is the first attempt to propose a unified setting named temporal event forecasting, considering both query and product oriented prediction in a temporal manner.
    \item \textbf{Novel framework}: We propose a novel RETE framework to enhance the  temporal event prediction for low-data users and meanwhile alleviate the over-smoothing issue by dynamically retrieving user-centered entities from the highly-connected query product evolutionary graph. 
    \item \textbf{Extensive evaluations}: Empirically, extensive experiments on both the public dataset and large-scale industrial datasets demonstrate the effectiveness of the proposed RETE method.
\end{itemize}

\section{related work}
\label{sec:related}

\begin{table}[t]
\caption{Comparison with existing settings. Action history refers to various past users' behaviors towards products such as ``Add'', ``Click'' and ``Purchase'', etc. Search history denotes historical user search queries. Meta data refers to product side information like attributes.}
\label{tb:setting}
\centering
\footnotesize
\fontsize{10}{10}\selectfont
\resizebox{1.0\columnwidth}{!}
{
\begin{tabular}{c|c|ccccc}
\hline
\multicolumn{2}{c|}{\textbf{Properties}} & \textbf{\begin{tabular}[c]{@{}c@{}}Action\\ history\end{tabular}} & \textbf{\begin{tabular}[c]{@{}c@{}}Search\\ history\end{tabular}} & \textbf{\begin{tabular}[c]{@{}c@{}}Meta\\ data\end{tabular}} & \textbf{\begin{tabular}[c]{@{}c@{}}Temporal\\ Info.\end{tabular}} & \textbf{\begin{tabular}[c]{@{}c@{}}Multi\\ objective\end{tabular}} \\ \hline
\multirow{3}{*}[-11pt]{Recommendation} & \begin{tabular}[c]{@{}c@{}}FM-based\\ \cite{FM,NFM}\end{tabular} & \cmark & \xmark & \cmark & \xmark & Product \\ \cline{2-2}
 & \begin{tabular}[c]{@{}c@{}}Sequential /\\ Session-based\\ \cite{BERT4Rec,GRU4Rec,userRNN,contextRNN,DySession}\end{tabular} &  \cmark & \xmark & \xmark & \cmark & Product  \\ \cline{2-2}
 & \begin{tabular}[c]{@{}c@{}}KG-based\\ \cite{KGAT,ECFKG,CKE,RippleNet}\end{tabular} & \cmark & \xmark & \cmark & \xmark & Product \\ \hline
 \multicolumn{2}{c|}{\begin{tabular}[c]{@{}c@{}}Query suggestion\\ \cite{query1,query2}\end{tabular}} & \xmark & \cmark & \xmark & \xmark & Query  \\ \cline{1-2}
\multicolumn{2}{c|}{\begin{tabular}[c]{@{}c@{}}Dynamic Graph Learning\\\cite{JODIE,CoEol}\end{tabular}} & \cmark & \xmark & \xmark & \cmark & Product  \\ \cline{1-2}

\multicolumn{2}{c|}{\begin{tabular}[c]{@{}c@{}}Temporal Event Forecasting\\(\textbf{Ours})\end{tabular}} & \cmark & \cmark & \cmark & \cmark & Product \& Query  \\ \hline
\end{tabular}
}
\vspace{-5mm}
\end{table}
We discuss and compare existing works from three related areas, as summarized in Table~\ref{tb:setting}.

\noindent\textbf{Product recommendation}.
Recommendation systems (RS) have achieved huge success on E-commerce platforms. Numerous efforts are devoted to improve RS from different perspectives: factorization machine (FM-based) models~\cite{FM,NFM,FM3} efficiently consider abundant input features, sequential/session-based recommendation models~\cite{BERT4Rec,GRU4Rec,userRNN,contextRNN,DySession,seqRec1} capture user-side dynamics from long/short time periods respectively, and knowledge graph (KG-based)~\cite{KGAT,ECFKG,RippleNet} models consider higher-order connections in multi-relational graph and produce explainable prediction. However, existing works merely focus on product-side interactions (e.g, user-to-product~\cite{FM,NFM,CoEol,JODIE}, product-to-product~\cite{BERT4Rec,GRU4Rec,P-Companion}, or product-to-attribute~\cite{KGAT,ECFKG,CKE,RippleNet}) but neglect equally important search queries. We study a more generic temporal event forecasting task to jointly optimize product and query prediction on evolutionary knowledge. A few interactive recommendation works on knowledge graph optimize sequential policy for recommending products within short sessions~\cite{IRS}. In contrast, we focus our attention on studying how to better integrate time-aware information from evolutionary knowledge graph and exclude unrelated noises in order to improve long-term performance for both product prediction and query prediction. 

\noindent\textbf{Query suggestion}.
Query suggestion task aims to assistant users to better express their intents on various search engine systems. General query suggestion task includes three different objectives: query rewriting (QR)~\cite{QR1,QR3}, query auto-completion (QAC)~\cite{QAC1,QAC2} and query prediction~\cite{query1,query2}. While QR and QAC focus on reformulating the queries that help the users to find better search results of current search intents, we specifically study query prediction that "recommend" queries that potentially match user intents. Existing works on recommending/predicting queries utilize search log data~\cite{query2}, query dependency graph~\cite{query3} or interaction history with users~\cite{query1}. Instead, to the best of our knowledge, we are the first to propose and study joint product and query prediction task on E-commerce. Inspired by the recent success of combining both queries and documents for jointly information retrieval in search engine systems~\cite{JointIR},  we mainly focus on exploring mutual benefits from both queries and products. One major difference is that we only consider structured graph data, instead of content, as query contents are much more private and sensitive on E-commerce.

\noindent\textbf{Dynamic graph learning}.
Graph learning methods~\cite{deepwalk,gcn,DGCN,hypergraph,bright,jing2021hdmi} has been widely explored for recommendation~\cite{KGAT,lightGCN}, user modeling~\cite{user2,user3,polarization,misinfo,DKGA}, etc. Recently, dynamic algorithms are proposed to better capture the temporal patterns of evolutionary graph~\cite{DySAT,DGCN,JODIE,DyDiff-VAE,network3}. As the constructed query product evolutionary graphs are more diverse and complicated, it brings new challenges to deal with over-smoothing issues and to capture temporal behavior patterns at the same time. Hence we aim to design a dynamic model with subgraph samplers to better learn informative and time-aware user representations. Although existing works have explored combinational power of subgraph samplers with graph neural networks on static and homogeneous/bipartite graph~\cite{shaDow,IMP}, we further facilitate dynamic graph learning with subgraph samplers on the evolutionary knowledge graph.
\section{Preliminaries and analysis}
\label{sec:pre}
\subsection{Concepts and Notations}
We study query-centric events and product-centric events associated with temporal information, i.e., a user $u$ types a search query $q$ or performs actions on a product $p$ at the time $t$. A set of event records constitutes to the dynamic interaction graph $\mathcal{G}_A^t = \{(e, r, e^{\prime}, t)| e\in \mathcal{U}, e^{\prime} \in \mathcal{Q} \bigcup \mathcal{P}, r \in \mathcal{R}_A\}$, where $\mathcal{U}$, $\mathcal{Q}$ and $\mathcal{P}$ denote the user set, query set and product set, respectively. Relation set $\mathcal{R}_A$ represents the interaction types among them, such as {\it ``type the query''}, {\it ``click the product''}, {\it ``adding product to carts''} and so on.

At the same time, rich meta-data of products forms a heterogeneous product graph $\mathcal{G}_P$, describing important attributes of each product $p \in \mathcal{P}$. Specifically, $\mathcal{G}_P = \{(e, r, e^{\prime})| e \in \mathcal{P}, e^{\prime} \in \mathcal{I} \bigcup \mathcal{Q}, r \in \mathcal{R}_P\}$, where $\mathcal{I}$ denotes attribute set for products, including but not limited to brand, product type and category. $\mathcal{R}_P$ denotes the relation set among them. $\mathcal{G}_P$ also describes match relations between products and queries.

\begin{figure}[t]
    \centering
    \includegraphics[width = 0.85\linewidth]{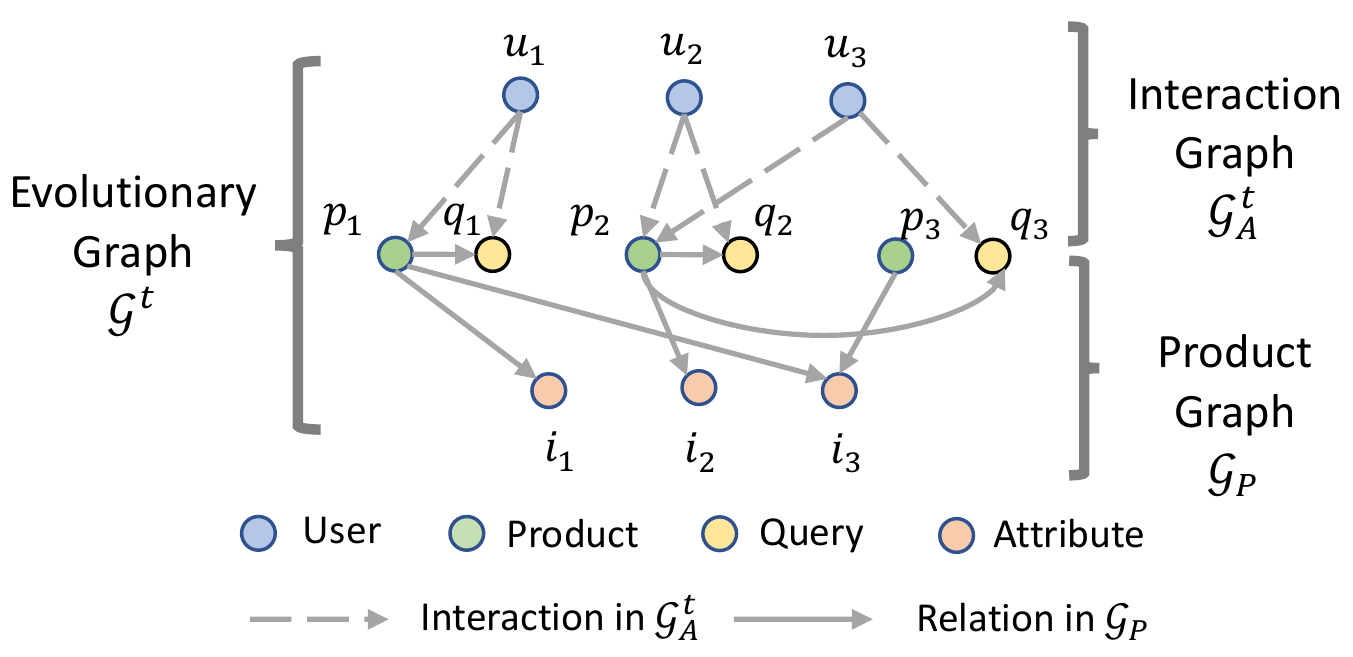}
    \caption{Query Product Evolutionary Graph at the $t$-th time.}
    \label{fig:graph}
    \vspace{-5mm}
\end{figure}

To capture the temporal patterns of user behaviors, we split time period into discrete time steps $t = 1, 2, \cdots, T$. Within each time step, for convenience of discussion, we unify the dynamic interaction graph $\mathcal{G}_A^t$ and the product graph $\mathcal{G}_P$ as a snapshot of \textit{evolutionary knowledge graph}: $\mathcal{G}^t = \mathcal{G}_A^t \bigcup \mathcal{G}_P$, as shown in Figure~\ref{fig:graph}.

\subsection{Definition}

\begin{definition}[\textbf{Temporal event forecasting}]
Given a collected evolutionary knowledge graph $\mathbb{G} = \{\mathcal{G}^1, \mathcal{G}^2, \cdots, \mathcal{G}^T\}$, for each user $u\in \mathcal{U}$, temporal event forecasting aims to predict potentially interacted query $q \in \mathcal{Q}$ and product $p \in \mathcal{P}$ after $T$.
\end{definition}

\subsection{Analysis}
\label{sec:analysis}
On the evolutionary knowledge graph, user intents can be captured by integrating abundant information from connected entities~\cite{CKE}. Existing KG-based frameworks~\cite{KGAT,RippleNet,ECFKG} map entities into a low-dimensional space, such that the relevance between user, query or product can be modeled via corresponding representations, i.e., $\boldsymbol{h}_u, \boldsymbol{h}_p, \boldsymbol{h}_q \in \mathbb{R}^d$. They propose various propagation mechanisms~\cite{gcn,GAT} on whole KG to integrate abundant information for each user, which can be generally described as follows:
\begin{equation}
    \boldsymbol{h}_u^{(l)} = \sum_{e^{\prime} \in \mathcal{N}_u} \pi (u, r, e^\prime) \boldsymbol{h}_{e^\prime}^{(l-1)},
    \label{eq:KG}
\end{equation}
\noindent
where $\mathcal{N}_u$ denotes neighbor set of user $u$, $l$ denotes the number of propagation layers, triple $(u, r, e^\prime)$ describes interaction between $u$ and $e^\prime$, and $\pi (u, r, e^\prime)$ denotes aggregation weights. 

Directly generalizing this family of propagation mechanisms to the event forecasting task, especially on evolutionary knowledge graph, faces two issues: (i) As $l$ grows to integrate higher-order information, the neighborhood size increases exponentially. A large ratio of unrelated entities (noises) are integrated, making user representations less distinguishable from each other, or even leading to over-smoothing~\cite{oversmoothing1,oversmoothing2,IMP}; (ii) evolutionary knowledge graph shows different distribution over time as users have evolving intents and behavior patterns. Ignoring such temporal factors not only fails to capture the most recent data characteristics but also worsens the first issue due to integrating a large ratio of out-of-fashion records for learning the representations.

\section{methodology}
\label{sec:model}

\begin{figure*}[t]
    \centering
    \includegraphics[width = 0.88\linewidth]{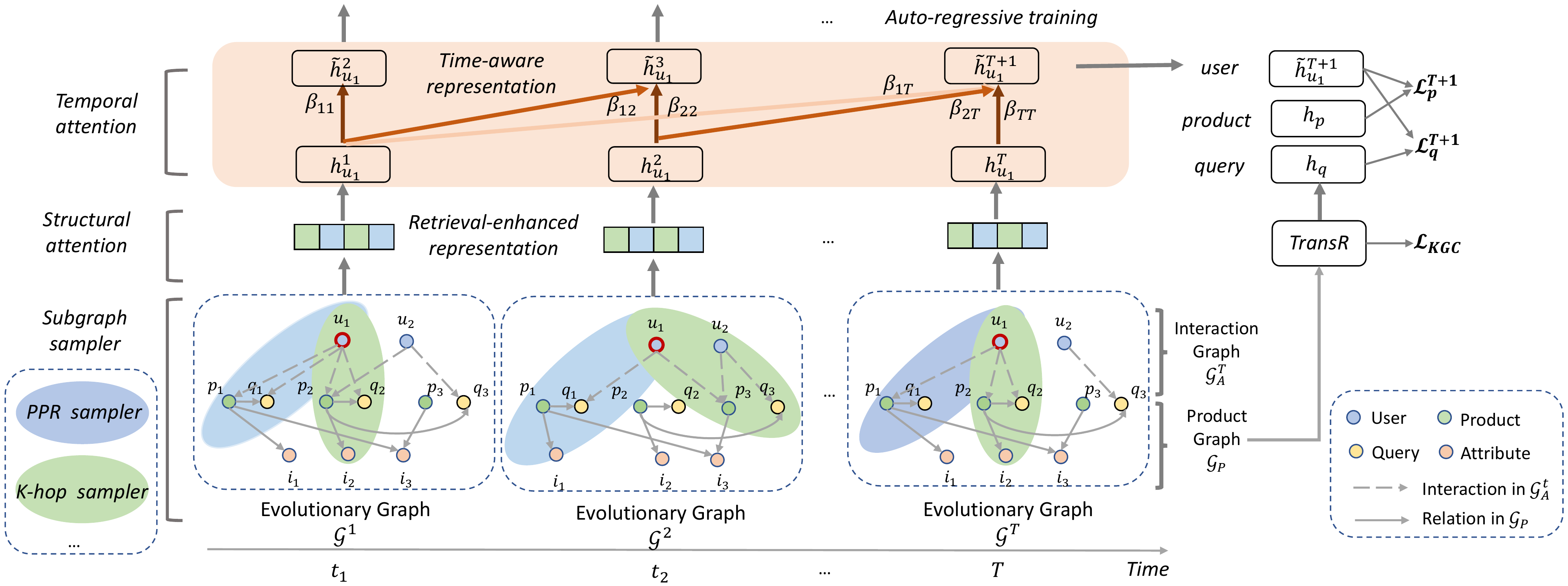}
    \vspace{-3mm}
    \caption{The framework of the proposed RETE model. We optimize the RETE model using the auto-regressive ranking loss and the knowledge graph completion loss (TransR) alternatively.}
    \label{fig:framework}
    \vspace{-3mm}
\end{figure*}

In this section, we present the proposed RETE framework. We firstly start with the overview of RETE and then detail three major components. Finally, we describe the model optimization.

\subsection{Overview}
\label{sec:requirement}
To address two issues mentioned in Section~\ref{sec:analysis} simultaneously, the proposed RETE strives to satisfy the following requirements:

\begin{itemize}[leftmargin = 15pt]
    \item \textit{Requirement 1}: It should learn informative and discriminative user representations by considering higher-order information and filtering out large ratio of noise from the graph.
    \item \textit{Requirement 2}: It should capture user intent evolution from data at different time steps, so as to produce up-to-date forecasting.
\end{itemize}

The framework of RETE is shown in Figure~\ref{fig:framework}. Based on the collected evolutionary knowledge graph, the key idea of RETE is to learn informative user representations at each time step from the sampled subgraphs instead of the whole graph, and combine them together via learnable temporal weights to capture user intent evolution. RETE first utilizes subgraph samplers to retrieve related entities centrally around each user $u$. And then it integrates rich information from the subgraphs via a structural attention module. A sequence of learned representations from all time steps are integrated via a temporal attention module. It can automatically learn the combination weights in the temporal domain so that more related (e.g., more recent) time steps are assigned larger weights, and unrelated (e.g., far away) time steps are assigned smaller weights (but not necessary to 0, as they may also reflect long-term intents).

To fulfill \textit{Requirement 1}, the sampled subgraphs constrain the attentive information propagation within local highly-related structure instead of the whole graph, so as to consider higher-order information and filter out noise. The design of the temporal module enables us to auto-regressively learn users' up-to-date representations to meet \textit{Requirement 2}. We refer readers to Appendix~\ref{ap:proof} for theoretical analysis of how RETE satisfies two requirements. Next, we will introduce the subgraph samplers, the structural attention module, and the temporal attention module respectively. 

\subsection{Ensemble subgraph sampler}
The proposed sampler aims to retrieve diverse and related entities via higher-order connections and excludes unrelated noises as order increases. To design strong indicators for such desirable subgraphs, we adopt structure-dependent Personalized PageRank (PPR) value~\cite{PPR1} which has been recently shown effective to retrieve related nodes from homogeneous graph~\cite{PPR_graph1,PPR_graph2,shaDow}. And we further ensemble it with a simple randomized $k$-hop sampler to retrieve entities more comprehensively. Advantages of this design, instead of utilizing trainable policy, are that it does not require reliable input features (unlike~\cite{IMP}, it is fully feature-independent) and it achieves much higher sample efficiency (almost real-time with careful implementation, as shown in Appendix~\ref{ap:implementation}). We summarize the designed sampler as follows:

\begin{itemize}[leftmargin = 15pt]
    \item \textbf{PPR sampler}. We use the feature-independent Personalized PageRank (PPR) value. Given a target user $u$, our PPR sampler first computes the approximate PPR value for all other entities, then selects up to $b$ neighborhood above threshold $\theta$ and preserves relations among selected entity set.
    \item \textbf{$k$-hop sampler}. Starting from a target user $u$, the $k$-hop sampler traverses up to $k$-hop connections and randomly selected up to $b$ neighbors. 
    \item \textbf{Ensemble sampler}. To capture a full picture of user intents, we ensemble multiple samplers under different types or with different parameters to parallelly sample several subgraphs.
\end{itemize}

\subsection{Structural attention module}
\begin{figure}[t]
    \centering
    \includegraphics[width = 0.85\linewidth]{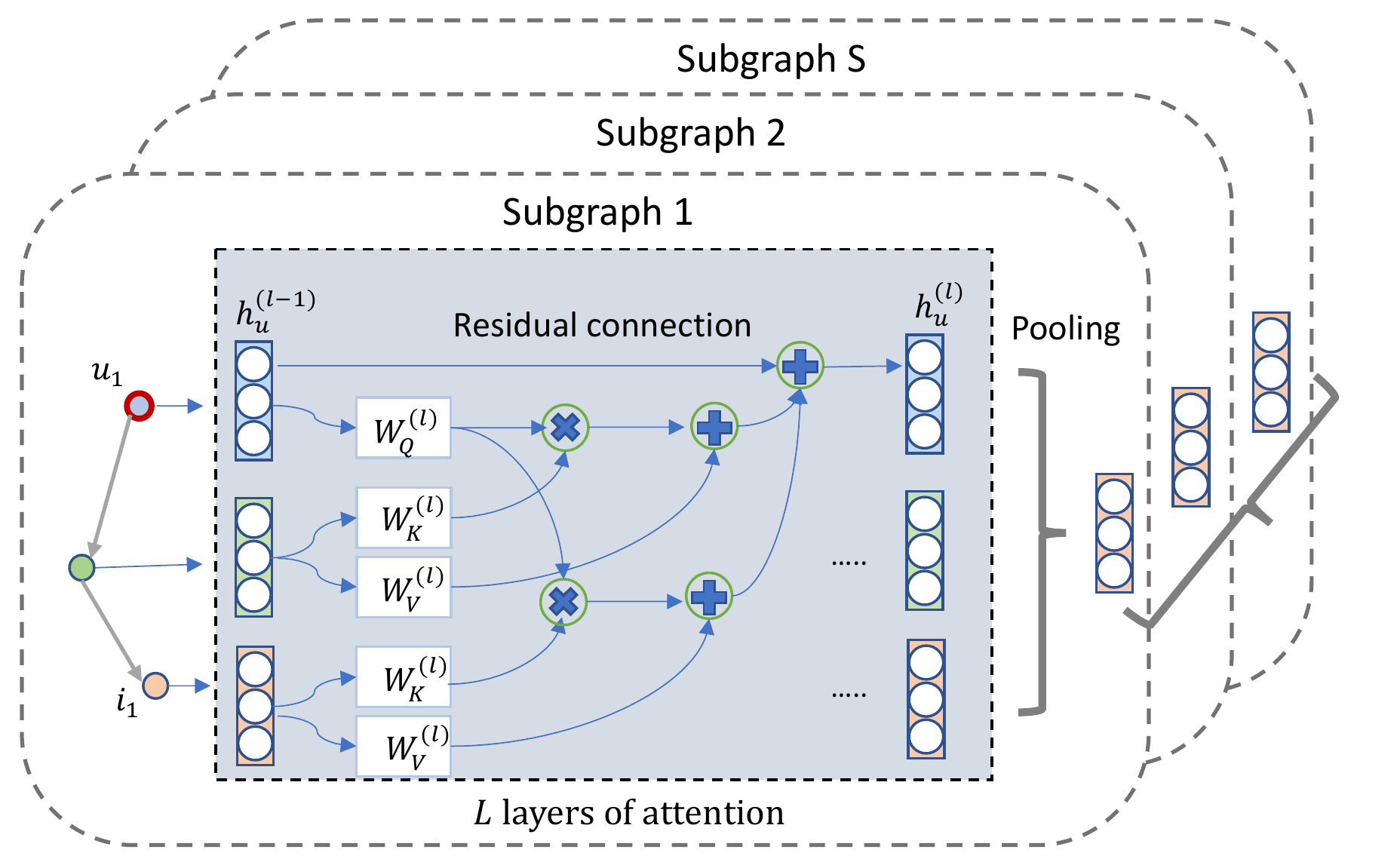}
    \caption{Structural attention module on subgraphs.}
    \label{fig:structural}
    \vspace{-4mm}
\end{figure}

Without loss of generality, let $s$ denote the number of subgraphs from the ensemble sampler. At time step $t$, given the sampled subgraphs $\{\mathcal{G}_{[u]}^{1t}, \cdots, \mathcal{G}_{[u]}^{st}\}$ for each user $u$, the structural attention module aims to integrate all useful information to learn user representations. As shown in Figure~\ref{fig:structural}, it first extracts information from each subgraph via multi-layer graph attentions~\cite{GAT}. Then to combine information from different perspectives, it fuses outputs from different subgraphs together to capture the global picture of user intent.

Specifically,  we first utilize $L$ graph attention layers to propagate and integrate information within each subgraph $\mathcal{G}_{[u]}^{st}$. Each layer can be summarized in Eq.~\ref{eq:structure1}:

\begin{equation}
    \boldsymbol{h}_{u}^{(l)}=\sigma\left(\sum_{v \in \mathcal{N}_{u}} \alpha_{uv}^{(l)} \boldsymbol{W}_{V}^{(l)} \boldsymbol{h}_{v}^{(l-1)}\right), (1 \leq l \leq L)
    \label{eq:structure1}
\end{equation}
\noindent
where $\boldsymbol{h}_{u}^{(l)}$ denotes user representations from layer $l$, $\mathcal{N}_{u}$ denotes neighbor set around user $u$ on subgraph $\mathcal{G}_{[u]}^{st}$, and $ \alpha_{uv}^{(l)}$ is aggregation attention weight shown in following equation:

\begin{equation}
    \alpha_{uv}^{(l)}=\frac{\exp \left(\sigma\left(\boldsymbol{a}^{T}\left[\boldsymbol{W}_{Q}^{(l)} \boldsymbol{h}_{u}^{(l)} \| \boldsymbol{W}_{K}^{(l)} \boldsymbol{h}_{v}^{(l)}\right]\right)\right)}{\sum_{v^\prime \in \mathcal{N}_{u}} \exp \left(\sigma\left(\boldsymbol{a}^{T}\left[\boldsymbol{W}_{Q}^{(l)} \boldsymbol{h}_{u}^{(l)} \| \boldsymbol{W}_{K}^{(l)} \boldsymbol{h}_{v^\prime}^{(l)}\right]\right)\right)},
    \label{eq:structure2}
\end{equation}
\noindent
where $\boldsymbol{W}_{V}^{(l)}$, $\boldsymbol{W}_{K}^{(l)}$, $\boldsymbol{W}_{Q}^{(l)}$ are shared weighted transformation applied to each entity in the subgraph, $\boldsymbol{a}$ is a weight vector parameterizing the attention function implemented as feed-forward layer, $\|$ is the concatenation operation and $\sigma(\cdot)$ is non-linear activation function. 

Since information propagation is constrained within the sampled subgraphs, more attention layers can be stacked to better learn latent representations without introducing other unrelated entities and noises. We then use residual connection across $L$ layers and graph-level pooling to better integrate information to capture user's intent at time step $t$:

\begin{equation}
    \boldsymbol{h}_u^{st} = \frac{1}{|\mathcal{E}_{[u]}^{st}| \times L} \sum_{e\in\mathcal{E}_{[u]}^{st}} \sum_{l} \boldsymbol{h}_{u}^{(l)},
\end{equation}
\noindent
where $\mathcal{E}_{[u]}^{st}$ denotes entity set in $s$-th subgraph $\mathcal{G}_{[u]}^{st}$. By doing so, we are able to integrate all useful information from the subgraphs to better represent users. To further capture a global view of user intent, ensemble samplers sample several subgraphs. From the learned representations of those subgraphs, we utilize one-layer MLP to integrate global information:

\begin{equation}
    \boldsymbol{h}_u^{t}= \sigma\left(\boldsymbol{W} \left[\boldsymbol{h}_u^{1t} \| \boldsymbol{h}_u^{2t} \| \cdots \| \boldsymbol{h}_u^{st}\right]\right).
    \label{eq:user}
\end{equation}

\subsection{Temporal attention module}
Given the learned user representations $\boldsymbol{H} = \{\boldsymbol{h}_u^{1}, \boldsymbol{h}_u^{2}, \cdots, \boldsymbol{h}_u^{T}\}$ from all time steps, we infer users' intents in a near future, i.e., $\widetilde{\boldsymbol{h}}_u^{T+1}$, for temporal event forecasting. We propose a temporal attention module to automatically capture both long-term and short-term intents by assigning different weights among $h_u^{t}$. The updating function for most recent user representation $\boldsymbol{h}_u^{T+1}$ can be expressed as follows:
\begin{equation}
    \widetilde{\boldsymbol{h}}_u^{T+1} = \sum_{t = 1}^{T} \beta_{tT} \boldsymbol{h}_u^{t}\boldsymbol{W}^V,
    \label{eq:temporal}
\end{equation}
\begin{equation}
    \beta_{ij} = \operatorname{softmax}\left(\frac{\boldsymbol{H} \boldsymbol{W}^Q(\boldsymbol{H} \boldsymbol{W}^K)^T}{\sqrt{d}} + \boldsymbol{M}\right)_{ij},
\end{equation}

\noindent
where $\boldsymbol{W}^Q$, $\boldsymbol{W}^L$, $\boldsymbol{W}^V$ are trainable temporal parameters, $\beta_{ij}$ is learned temporal weight, $d$ denotes dimension of user representations, and $\boldsymbol{M}$ is added to ensure auto-regressive setting, i.e., preventing future information affecting current state. We define $\boldsymbol{M}_{ij}=0$ if $i\leq j$, otherwise $-\infty$.

By applying Eq.~\ref{eq:temporal}, we are able to not only emphasize those information related to users' short-term intents represented in $\boldsymbol{h}_u^{T}$, but also capture long-term intents as we integrate information from all time steps. The proposed temporal module provides better interpretability, where the temporal attention weights reflect user intent evolution and shifting. Notably, it can be auto-regressively applied to newly collected data from $T+2, T+3, \cdots$, as it automatically computes temporal combination weights in the future time steps without model modification or retraining.

\subsection{Optimization}
The proposed framework is expected to capture the evolution of user preference from the evolutionary knowledge graph. To better represent product and query information from static product graph into the same low-dimensional space, we utilize a knowledge graph embedding module TransR~\cite{TransR} to optimize knowledge graph completion loss $\mathcal{L}_{KGC}$, which is detailed in Appendix~\ref{ap:transr}. After learning user/product/query representations at time step $t$, i.e., $\widetilde{\boldsymbol{h}}_u^t$, $\boldsymbol{h}_p$, $\boldsymbol{h}_q$, we use inner product $\gamma^t_{up} = <\widetilde{\boldsymbol{h}}_u^t, \boldsymbol{h}_p>$ and $\gamma^t_{uq} = <\widetilde{\boldsymbol{h}}_u^t, \boldsymbol{h}_q>$ to model relevance.

We use negative sampling to accelerate and stabilize the training process. At time $t$, for each user-product pair $(u,p^+)$, we randomly sample several negative samples $(u,p^-)$, where we expect $\gamma^t_{up-}$ is smaller than $\gamma^t_{up+}$ by a margin. Thus, we adopt the weighted approximate-rank pairwise (WARP) loss~\cite{warp} for product prediction as follows:
\begin{equation}
    \mathcal{L}_{p} = \sum_{t=1}^T\mathbb{E}_{(u, p^{+})\in \mathcal{G}^t} \sum_{p^{-}} \frac{L(rank(p^{+})) \cdot |\lambda_m - \gamma^t_{up+} + \gamma^t_{up-}|_{+}}{rank(p^{+})},
    \label{eq:diffloss}
\end{equation}
\noindent
where $\lambda_m$ denotes margin value, $|\cdot|_{+}$ means $\max(0,\cdot)$. For each observed interaction $(u, p^{+})$, we expect the relevance score $\gamma^t_{up+}$ to be larger than that of any negative samples by  $\lambda_m$, i.e., $|\lambda_m - \gamma^t_{up+} + \gamma^t_{up-}|_{+} = 0$. Otherwise, we penalize each pair of $(p^{+}, p^{-})$ because of the incorrect ranking. $L(K) = \sum_{k = 1}^{K}1/k$, $rank(p^{+})$ denotes relative ranking of positive sample $p^{+}$ among negative samples $p^{-}$, and $L(rank(p^{+}))$ is the penalty weight. Similarly we can define WARP loss for query prediction as $\mathcal{L}_{q}$. Thus, the overall objective is:
\begin{equation}
\mathcal{L}=\mathcal{L}_{p}+\mathcal{L}_{q}+\mathcal{L}_{KGC}+\| \Theta\|_{2},
\end{equation}
\noindent
where $\Theta$ denotes model parameters, $\mathcal{L}_{KGC}$ denotes TransR loss to represent static facts in the product graph. We optimize $\mathcal{L}_{\operatorname{KGC}}$ and $\mathcal{L}_{p}+\mathcal{L}_{q}+\| \Theta\|_{2}$ alternatively, where mini-batch Adam is adopted. We summarize the optimization process of RETE in Algorithm~\ref{al:training}.

\begin{algorithm}[t]
\caption{The optimization process for RETE.}
\label{al:training}
\KwIn{Evolutionary knowledge graph $\{\mathcal{G}^1, \cdots, \mathcal{G}^T\}$ and product graph $\mathcal{G}_P$.}
\KwOut{User intents $\widetilde{\boldsymbol{h}}^{T+1}$ and model parameters $\Theta$.}
Ensemble samplers sample subgraphs for users; \\
\While{model not converged}{
    \textbf{Optimize on product graph}:\\
    Minimize $\mathcal{L}_{KGC}$ and update entity representation in $\mathcal{G}_P$; \\
    \textbf{Optimize on evolutionary graph}: \\
    \For{Each time step $t < T$ during training}{
        \textit{(Auto-regressive training:)}\\
        Learn user intent $\{\boldsymbol{h}^{1}, \cdots, \boldsymbol{h}^{t-1}\}$ according to Eq.~\ref{eq:user};\\
        Update new intents $\widetilde{\boldsymbol{h}}^{t}$ according to Eq.~\ref{eq:temporal}; \\
        Calculate ranking loss $\mathcal{L}^t_p$ and $\mathcal{L}^t_q$ at time step $t$;\\
    }
    Optimizing ranking loss $\mathcal{L}_p$ and $\mathcal{L}_q$ in Eq.~\ref{eq:diffloss};
}
Update user intents $\widetilde{\boldsymbol{h}}^{T+1}$ according to Eq.~\ref{eq:temporal};
\end{algorithm}


\section{experiment}
\label{sec:experiment}

We evaluate RETE on one public and four real-world E-commerce datasets, and we aim to answer the following research questions:
\begin{itemize}[leftmargin = 15pt]
    \item \textbf{RQ1}: How does RETE perform compared with state-of-the-art models on the datasets in both academia and industry?
    \item \textbf{RQ2}: How do different components affect RETE performance?
    \item \textbf{RQ3}: Can RETE better integrate information from neighbors?
    \item \textbf{RQ4}: Can RETE capture the evolution of users’ preferences?
\end{itemize}

\subsection{Experimental setup}

\subsubsection{Datasets}
We collected one public Yelp dataset and four industrial E-commerce datasets for experiments:

\begin{itemize}[leftmargin = 15pt]
    \item \textbf{Yelp}. The dataset is adopted in Yelp Challenge 2019 \footnote{https://www.yelp.com/dataset/}, which contains the interaction records between users and businesses like restaurants and bars. For ease of evaluation, we extract data since April 2014, spanning a period of more than $~\sim7$ years. We generate pseudo queries by extracting representitive keyphrases from user reviews. And we remove users, products, queries with less interactions than $20$. To construct product graph we use attributes like category, location, etc. 
    \item \textbf{E-commerce}. We gain access to the search log data spanning a period of $140$ days and product attribute data. We first collect data under four specific categories: \textit{Electronics}, \textit{Book}, \textit{Music} and \textit{Beauty}. For each categories, we retain users, products, queries with at least $10$ interactions. Then to construct product graph, we preserve product attributes, including brand, product type, etc. 
\end{itemize}

 We purposefully choose the two platform because of various length of time range. To evaluate our framework, we divide time span into $28$ time steps according to interaction timestamps. We split them into background/training/val/test (10/10/2/6) to train initial entity embeddings (model input), train, validate and test RETE respectively. We also try different time segmentation strategies, where our method consistently outperforms others. We leave a systematic study for optimal segmentation as our future work. Table~\ref{tb:data} in Appendix~\ref{ap:setup} summarizes the statistics of the experimental datasets.

\begin{table*}[t]
\caption{Overall performance for product and query prediction. Average results on $5$ independent runs are reported. $*$ indicates the statistically significant improvements over the best baseline, with $p$-value smaller than $0.001$.}
\label{tb:performance}
\small
\begin{subtable}{0.93\textwidth}
\centering
\resizebox{0.93\columnwidth}{!}{
\begin{tabular}{c|cc|cccccccc}
  \toprule
\multirow{2}{*}{{\textbf{Dataset}}} & \multicolumn{2}{c|}{\textbf{Public}} & \multicolumn{8}{c}{\textbf{Industrial E-commerce}} \\ \cline{2-11}
 & \multicolumn{2}{c|}{\textbf{Yelp}} & \multicolumn{2}{c}{\textbf{Electronics}} & \multicolumn{2}{c}{\textbf{Music}} & \multicolumn{2}{c}{\textbf{Book}} & \multicolumn{2}{c}{\textbf{Beauty}} \\ \hline
{$K = 20$} & {NDCG@$K$} & {Recall@$K$} & {NDCG@$K$} & \multicolumn{1}{c}{{Recall@$K$}} & {NDCG@$K$} & \multicolumn{1}{c}{{Recall@$K$}} & {NDCG@$K$} & \multicolumn{1}{c}{{Recall@$K$}} & {NDCG@$K$} & {Recall@$K$} \\ \hline
\multicolumn{11}{c}{ \textbf{FM-based Recommendation}} \\ \hline
{FM} & 0.0221 & 0.0277 & 0.0512 & \multicolumn{1}{c}{0.0713} & 0.0641 & \multicolumn{1}{c}{0.0981} & 0.0682 & \multicolumn{1}{c}{0.0964} & 0.1155 & 0.1459 \\   
{NFM} & 0.0214 & 0.0281 & 0.0715 & \multicolumn{1}{c}{0.1164} & 0.0761 & \multicolumn{1}{c}{0.1005} & 0.0793 & \multicolumn{1}{c}{0.1064} & 0.1246 & 0.1591 \\  \hline 
\multicolumn{11}{c}{ \textbf{Sequential Recommendation}} \\  \hline 
{BERT4Rec} & 0.0422 & 0.0501 & 0.0619 & \multicolumn{1}{c}{0.0832} & 0.0537 & \multicolumn{1}{c}{0.0618} & 0.0447 & \multicolumn{1}{c}{0.0651} & 0.0827 & 0.1015 \\
{GRU4Rec} & 0.0419 & 0.0511 & 0.0742 & \multicolumn{1}{c}{0.0859} & 0.0621 & \multicolumn{1}{c}{0.0711} & 0.0412 & \multicolumn{1}{c}{0.0658} & 0.0842 & 0.1003 \\ \hline
\multicolumn{11}{c}{ \textbf{Dynamic Graph Learning}} \\     \hline  

{JODIE} & 0.0459 & 0.0527 & 0.1399 & \multicolumn{1}{c}{0.1515} & 0.1123 & \multicolumn{1}{c}{0.1405} & 0.1401 & \multicolumn{1}{c}{0.1881} & 0.1458 & 0.1807 \\ \hline
\multicolumn{11}{c}{ \textbf{KG-based recommendation}} \\ \hline  
{KGAT} & 0.0342 & 0.0403 & 0.1503 & \multicolumn{1}{c}{0.1914} & 0.1156 & \multicolumn{1}{c}{0.1301} & 0.1254 & \multicolumn{1}{c}{0.1479} & 0.1503 & 0.1893 \\
{ECFKG} & 0.0388 & 0.0495 & 0.1413 & \multicolumn{1}{c}{0.1859} & 0.1036 & \multicolumn{1}{c}{0.1246} & 0.1327 & \multicolumn{1}{c}{0.1674} & 0.1401 & 0.1799 \\     \hline  
\textbf{RETE (Ours)} & \textbf{0.0499*} & \textbf{0.0589*} & \textbf{0.1703*} & \multicolumn{1}{c}{\textbf{0.2120*}} & \textbf{0.1304*} & \multicolumn{1}{c}{\textbf{0.1521*}} & \textbf{0.1455*} & \multicolumn{1}{c}{\textbf{0.1976*}} & \textbf{0.1621*} & \textbf{0.1985*} \\ \hline
\textit{{Gain}} & \textit{8.71\%} & \textit{11.76\%} & \textit{13.31\%} & \multicolumn{1}{c}{\textit{10.76\%}} & \textit{12.80\%} & \multicolumn{1}{c}{\textit{8.27\%}} & \textit{3.85\%} & \multicolumn{1}{c}{\textit{5.05\%}} & \textit{7.85\%} & \textit{4.86\%} \\     \bottomrule
\end{tabular}}
\caption{Product prediction performance.}
\label{tb:product}
\end{subtable}

\begin{subtable}{0.93\textwidth}
\centering
\resizebox{0.93\columnwidth}{!}{
\begin{tabular}{c|cc|cccccccc}
\toprule
\multirow{2}{*}{{\textbf{Dataset}}} & \multicolumn{2}{c|}{\textbf{Public}} & \multicolumn{8}{c}{\textbf{Industrial E-commerce}} \\ \cline{2-11}
 & \multicolumn{2}{c|}{\textbf{Yelp}} & \multicolumn{2}{c}{\textbf{Electronics}} & \multicolumn{2}{c}{\textbf{Music}} & \multicolumn{2}{c}{\textbf{Book}} & \multicolumn{2}{c}{\textbf{Beauty}} \\ \hline
{$K = 20$} & {NDCG@$K$} & {Recall@$K$} & {NDCG@$K$} & \multicolumn{1}{c}{{Recall@$K$}} & {NDCG@$K$} & \multicolumn{1}{c}{{Recall@$K$}} & {NDCG@$K$} & \multicolumn{1}{c}{{Recall@$K$}} & {NDCG@$K$} & {Recall@$K$} \\ \hline
\multicolumn{11}{c}{ \textbf{FM-based Recommendation}} \\ \hline
{FM} & 0.0257 & 0.0319 & 0.0481 & 0.0765 & 0.0324 & 0.0681 & 0.0862 & 0.1015 & 0.0614 & 0.0854 \\
{NFM} & 0.0244 & 0.0331 & 0.0533 & 0.0709 & 0.0583 & 0.1188 & 0.0851 & 0.1103 & 0.0673 & 0.0903 \\ \hline
\multicolumn{11}{c}{ \textbf{Sequential Recommendation}} \\  \hline 
{BERT4Rec} & 0.0407 & 0.0498 & 0.0602 & 0.0877 & 0.0207 & 0.0457 & 0.0413 & 0.0882 & 0.0417 & 0.0566 \\
{GRU4Rec} & 0.0381 & 0.0477 & 0.0590 & 0.0731 & 0.0436 & 0.0599 & 0.0401 & 0.0907 & 0.0513 & 0.0602 \\ \hline
\multicolumn{11}{c}{ \textbf{Dynamic Graph Learning}} \\     \hline 
{JODIE} & 0.0461 & 0.0617 & 0.0779 & 0.0957 & 0.0988 & 0.1364 & 0.1301 & 0.1475 & 0.1327 & 0.1495 \\ \hline
\multicolumn{11}{c}{ \textbf{KG-based recommendation}} \\ \hline 
{KGAT} & 0.0431 & 0.0527 & 0.0913 & 0.1153 & 0.0823 & 0.1324 & 0.1293 & 0.1497 & 0.1299 & 0.1502 \\
{ECFKG} & 0.0397 & 0.0481 & 0.0899 & 0.1099 & 0.0897 & 0.1259 & 0.1283 & 0.1503 & 0.1214 & 0.1518 \\ \hline
\textbf{RETE (Ours)} & \textbf{0.0507*} & \textbf{0.0653*} & \textbf{0.1015*} & \textbf{0.1393*} & \textbf{0.1033*} & \textbf{0.1408*} & \textbf{0.1391*} & \textbf{0.1557*} & \textbf{0.1487*} & \textbf{0.1643*} \\ \hline
\textit{{Gain}} & \textit{9.98\%} & \textit{5.83\%} & \textit{11.17\%} & \textit{20.82\%} & \textit{4.55\%} & \textit{3.22\%} & \textit{6.92\%} & \textit{4.01\%} & \textit{12.06\%} & \textit{9.31\%} \\ 
\bottomrule
\end{tabular}}
\caption{Query prediction performance.}
\label{tb:query}
\end{subtable}
\vspace{-5mm}
\end{table*}

\subsubsection{Metrics}
We evaluate temporal event forecasting task in a retrieval setting, i.e., we compare the predicted top-$K$ ranking list of products/queries with the groundtruth in the testing time steps. We adopt two widely-used evaluation protocols: \textit{Recall@$K$} and \textit{NDCG@$K$}. By default, we set $K = 20$.

\subsubsection{Baselines}
We compare baselines from following areas:
\begin{itemize}[leftmargin = 15pt]
    \item FM-based recommendation, which considers the second-order feature interactions. We compare \textbf{FM}~\cite{FM} and \textbf{NFM}~\cite{NFM}. 
    \item Sequential recommendation, which considers user evolving intents overtime. We compare \textbf{GRU4Rec}~\cite{GRU4Rec} and \textbf{BERT4Rec}~\cite{BERT4Rec}. 
    \item KG-based recommendation, which models heterogeneous entities and high-order connections for recommendation. We compare \textbf{ECFKG}~\cite{ECFKG} and \textbf{KGAT}~\cite{KGAT}.
    \item dynamic graph learning: which models evolutionary interaction graph. We compare \textbf{JODIE}~\cite{JODIE}.
\end{itemize}

Details can be found in Appendix~\ref{ap:setup}, including data collection, data statistics, baseline/model setup, hyper-parameter tuning, etc.

\subsection{Model Performance (RQ1)}
\subsubsection{Overall performance}

We first compare overall performance of product prediction and query prediction with selected baselines, as shown in  Table~\ref{tb:product} and Table~\ref{tb:query}. In most cases, FM-based (FM, NFM) and sequential recommendation (BERT4Rec, GRU4Rec) methods produce poor results, as they do not explicitly consider higher-order interactions. RETE beats KG-based methods (KGAT, ECFKG) and dynamic graph learning method (JODIE) on all metrics, as we propose a better way to integrate multi-relational data in a temporal manner. Notably, on E-commerce platform, query prediction has worse performance than predicting products, while Yelp platform exhibits different pattern. We hypothesis that it is because the real queries from users on E-commerce platform are more diverse than pseudo queries extracted from Yelp review data. Also, it is harder to produce accurate prediction on Yelp platform, as Yelp data are collected from much longer period, where user intent shifting and evolution across time step are much harder to capture.

\subsubsection{Detailed performance}
\begin{figure}[t]
    \centering
    \begin{subfigure}[b]{0.495\linewidth}
         \centering
         \includegraphics[width = 1.0 \linewidth, height = 0.71 \linewidth]{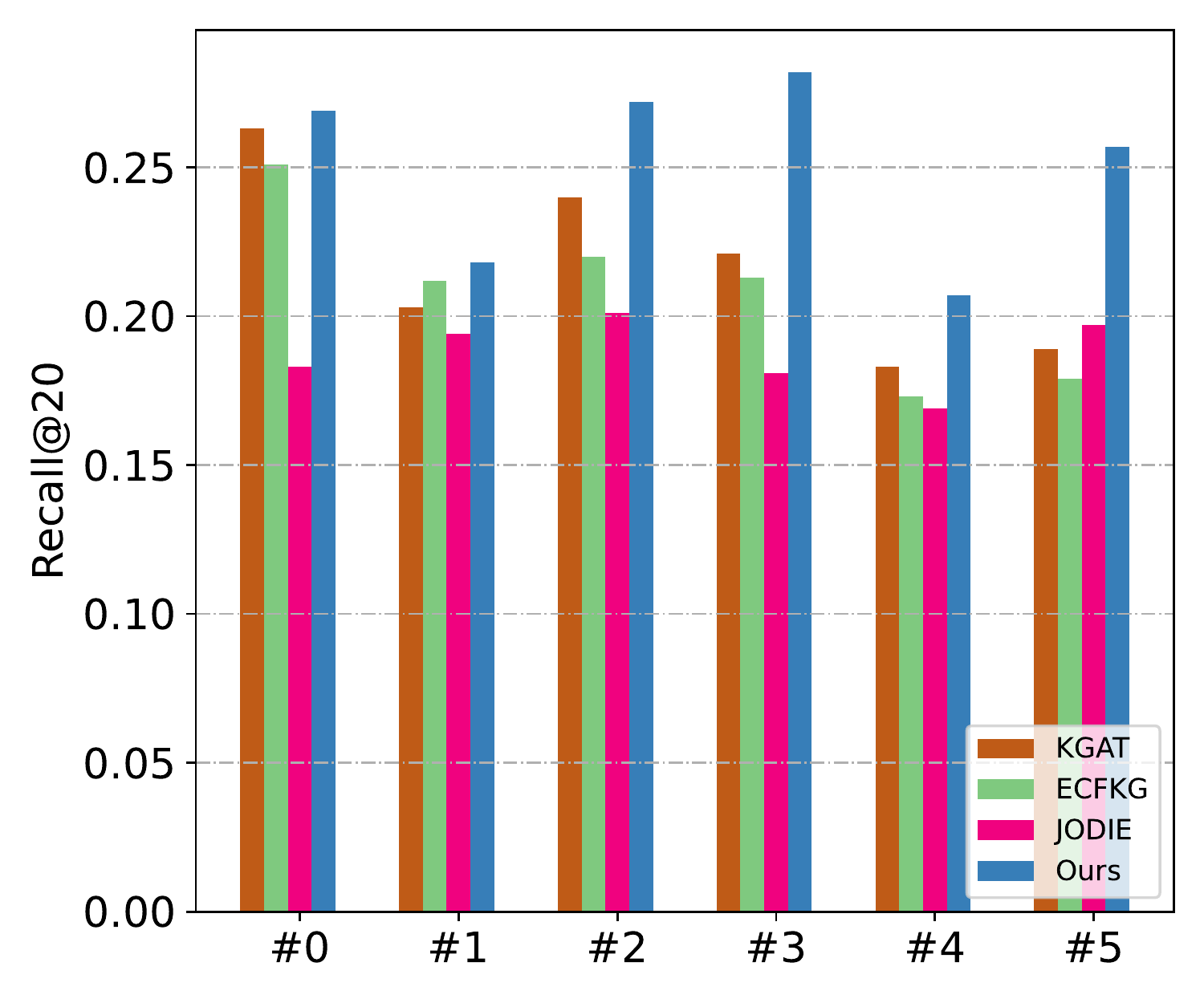}
         \caption{Product.}
         \label{fig:detailedP}
    \end{subfigure}
    \begin{subfigure}[b]{0.495\linewidth}
         \centering
         \includegraphics[width = 1.0 \linewidth, height = 0.71 \linewidth]{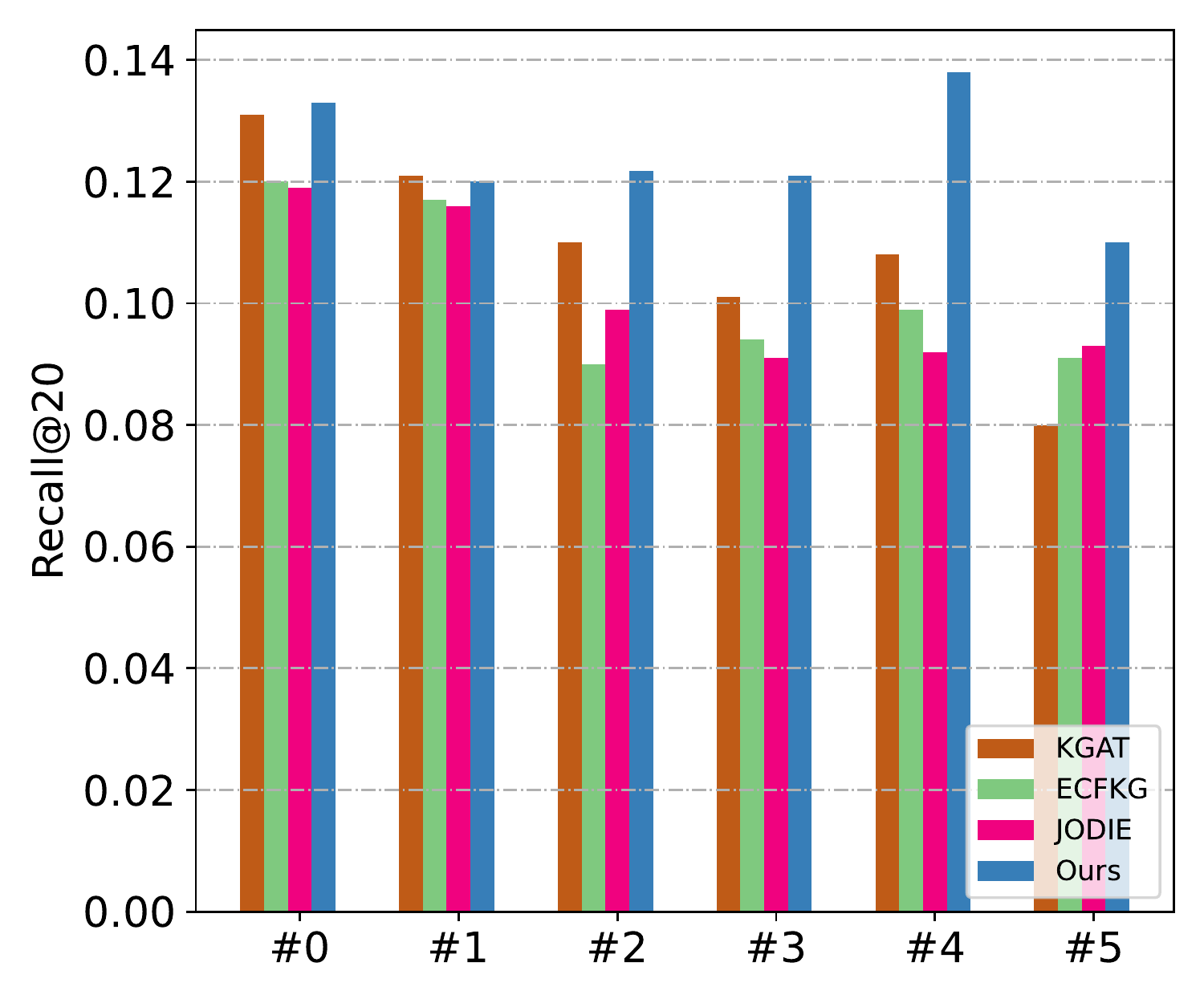}
         \caption{Query.}
         \label{fig:detailedQ}
    \end{subfigure}
    \caption{Recall@20 of each test period on \textit{Electronics}, significant improvements come from the last several time steps.}
    \label{fig:detail}
    \vspace{-5mm}
\end{figure}

Further, to investigate how does RETE perform over time, we compare the detailed performances in each testing time step ($6$ time step), as shown in Figure~\ref{fig:detail}. The performances in different time steps vary largely, indicating user intents are evolving and shifting. RETE can beat others in almost all time steps, and more significant improvements come from the last several time steps, which shows our proposed temporal module can capture the evolution of user preference and thus achieve better long-term performance.

\subsubsection{Auto-regressive evaluation.}
As users keep interacting with E-commerce platforms, new interaction events are collected continuously. In real scenario, it is required that the deployed models can take newly collected data to update user representations without time-consuming retraining or fine-tuning. We refer it to \textit{auto-regressive} evaluation and verify the robustness of RETE under it. As it is hard to update static models on new data without retraining, we mainly focus on comparing with dynamic models (BERT4Rec, GRU4Rec and JODIE). Given new testing data, we continuously fed them into the temporal module and evaluate the performance in the next time step.  Table~\ref{tb:auto-regressive} reports the average performance on testing time steps. All compared models achieve improved results after considering newly collected data, as which contain more up-to-date clues to capture users' intents. RETE can achieve the best performance, showing better generalization ability and robustness for continual learning.

\begin{table}[t]
\caption{Performance under the auto-regressive evaluation. Average results on six testing time steps are reported.}
\label{tb:auto-regressive}
\small
\resizebox{0.90\columnwidth}{!}{
\begin{tabular}{c|cccc}
\toprule
\textbf{Task} & \multicolumn{4}{c}{\textbf{Product prediction}} \\ \hline
\textbf{Dataset} & \multicolumn{2}{c|}{\textbf{Electronics}} & \multicolumn{2}{c}{\textbf{Music}} \\ \hline
\textbf{k = 20} & \textbf{NDCG@$K$} & \multicolumn{1}{c|}{\textbf{Recall@$K$}} & \textbf{NDCG@$K$} & \textbf{Recall@$K$} \\ \hline
\textbf{BERT4Rec} & 0.0830 & \multicolumn{1}{c|}{0.1232} & 0.0566 & 0.0701 \\
\textbf{GRU4Rec} & 0.1099 & \multicolumn{1}{c|}{0.1201} & 0.0519 & 0.0803 \\
\textbf{JODIE} & 0.1801 & \multicolumn{1}{c|}{0.1962} & 0.1371 & 0.1507 \\
\textbf{Ours} & \textbf{0.1961} & \multicolumn{1}{c|}{\textbf{0.2414}} & \textbf{0.1561} & \textbf{0.1733} \\ \hline \hline
\textbf{Task} & \multicolumn{4}{c}{\textbf{Query prediction}} \\ \hline 
\textbf{Dataset} & \multicolumn{2}{c|}{\textbf{Electronics}} & \multicolumn{2}{c}{\textbf{Music}} \\ \hline
\textbf{k = 20} & \textbf{NDCG@$K$} & \multicolumn{1}{c|}{\textbf{Recall@$K$}} & \textbf{NDCG@$K$} & \textbf{Recall@$K$} \\ \hline
\textbf{BERT4Rec} & 0.0861 & \multicolumn{1}{c|}{0.1019} & \multicolumn{1}{r}{0.0455} & \multicolumn{1}{r}{0.0634} \\
\textbf{GRU4Rec} & 0.0661 & \multicolumn{1}{c|}{0.0913} & \multicolumn{1}{r}{0.0501} & \multicolumn{1}{r}{0.0633} \\
\textbf{JODIE} & 0.1259 & \multicolumn{1}{c|}{0.1526} & \multicolumn{1}{r}{0.1203} & \multicolumn{1}{r}{0.1499} \\
\textbf{Ours} & \textbf{0.1425} & \multicolumn{1}{c|}{\textbf{0.1793}} & \multicolumn{1}{r}{\textbf{0.1352}} & \multicolumn{1}{r}{\textbf{0.1631}} \\ 
\bottomrule
\end{tabular}}
\vspace{-3mm}
\end{table}

\begin{table}[t]
\caption{Ablation study evaluated by Recall@$20$.}
\label{tb:ablation}
\small
\resizebox{0.9\columnwidth}{!}{
\begin{tabular}{ccccc}
\toprule
\multicolumn{1}{c|}{\textbf{Datasets}} & \multicolumn{2}{c|}{\textbf{Electronics}} & \multicolumn{2}{c}{\textbf{Music}} \\ \hline
\multicolumn{1}{c|}{\textbf{Ablations}} & \multicolumn{1}{c|}{\textbf{Product}} & \multicolumn{1}{c|}{\textbf{Query}} & \multicolumn{1}{c|}{\textbf{Product}} & \textbf{Query} \\ \hline\hline
\multicolumn{5}{l}{\textit{Variants on how to construct input graph:}} \\
\multicolumn{1}{c|}{\textbf{w/o attr.}} & 0.1686 & \multicolumn{1}{c|}{0.1037} & 0.1154 & 0.1132 \\
\multicolumn{1}{c|}{\textbf{w/o query}} & 0.1749 & \multicolumn{1}{c|}{-} & 0.1335 & - \\
\multicolumn{1}{c|}{\textbf{w/o product}} & - & \multicolumn{1}{c|}{0.0973} & - & 0.0943 \\ 
\hline\hline
\multicolumn{5}{l}{\textit{Variants on subgraph sampler:}} \\
\multicolumn{1}{c|}{\textbf{Only k-hop sampler}} & 0.1991 & \multicolumn{1}{c|}{0.1203} & 0.1363 & 0.1367 \\
\multicolumn{1}{c|}{\textbf{Only PPR sampler}} & 0.2123 & \multicolumn{1}{c|}{0.1381} & 0.1501 & 0.1399 \\
\hline\hline
\multicolumn{5}{l}{\textit{Static v.s. dynamic:}} \\
\multicolumn{1}{c|}{\textbf{Ours (static)}} & 0.1931 & \multicolumn{1}{c|}{0.1183} & 0.1299 & 0.1327 \\
\multicolumn{1}{c|}{\textbf{Ours}} & \textbf{0.2120} & \multicolumn{1}{c|}{\textbf{0.1393}} & \textbf{0.1521} & \textbf{0.1408} \\ 
\bottomrule
\end{tabular}}
\vspace{-3mm}
\end{table}

\subsection{Ablation Study (RQ2)}

To investigate how each component affects the model performance, we conduct the following ablation studies, as shown in Table~\ref{tb:ablation}:
\subsubsection{The effects of various types of information}: 
Our solution constructs a temporal KG to organize multi-relational data for joint product and query prediction. As expected, removing rich attributes causes a performance hit, because they provide reliable relations among products and queries. Furthermore, we can demonstrate the mutual benefit derived from the joint query and product prediction task via the performance degradation after removing product entities and query entities, respectively. Interestingly, removing products can significantly affect query prediction performance. This is because queries are mainly connected by product nodes, and a large ratio of query connections is ignored in this ablation.

\subsubsection{The effects of ensemble subgraph samplers}: 
We propose an ensemble subgraph sampler to retrieve relevant entities and filter unrelated noise from the whole graph. Different samplers can capture data characteristics from different perspectives. Only using the k-hop sampler or PPR sampler can hurt the performance compared with ensembling them together. The PPR sampler behaves better than the k-hop sampler, as PPR value can better reflect the relevance among entities when raw neighbor information.

\subsubsection{The effects of temporal module}: 
To evaluate the impact of the temporal attention module, we compare the performance of our static variant. Our dynamic model can have $\sim10\%$ relative improvements over the static variant.

\subsection{Analysis of ensemble sampler (RQ3)}
\begin{figure}[t]
    \centering
    \begin{subfigure}[b]{0.495\linewidth}
         \centering
         \includegraphics[width = \linewidth]{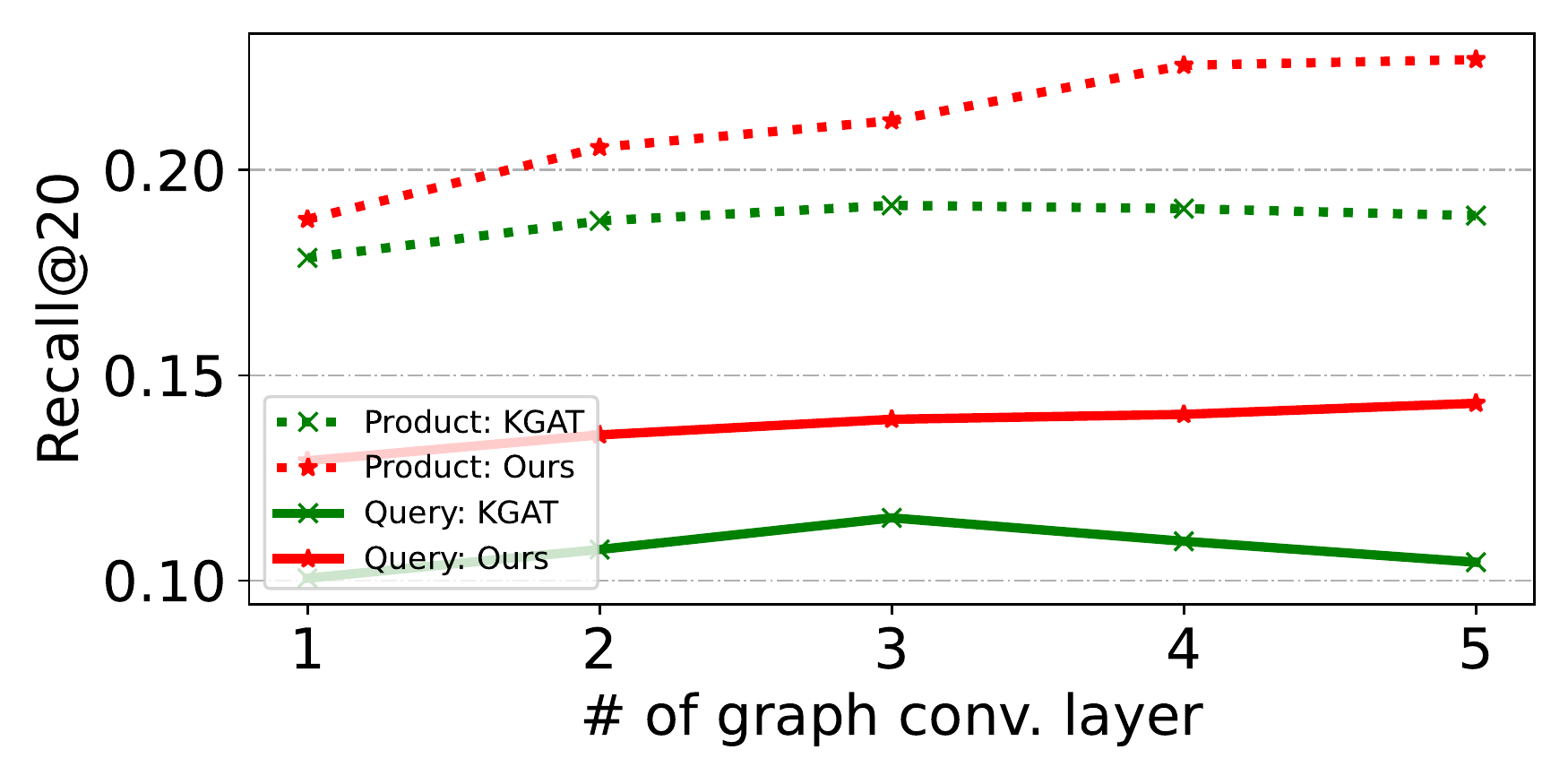}
    \caption{Effects of \# of layers.}
    \label{fig:layer}
    \end{subfigure}
    \begin{subfigure}[b]{0.495\linewidth}
         \centering
         \includegraphics[width = \linewidth]{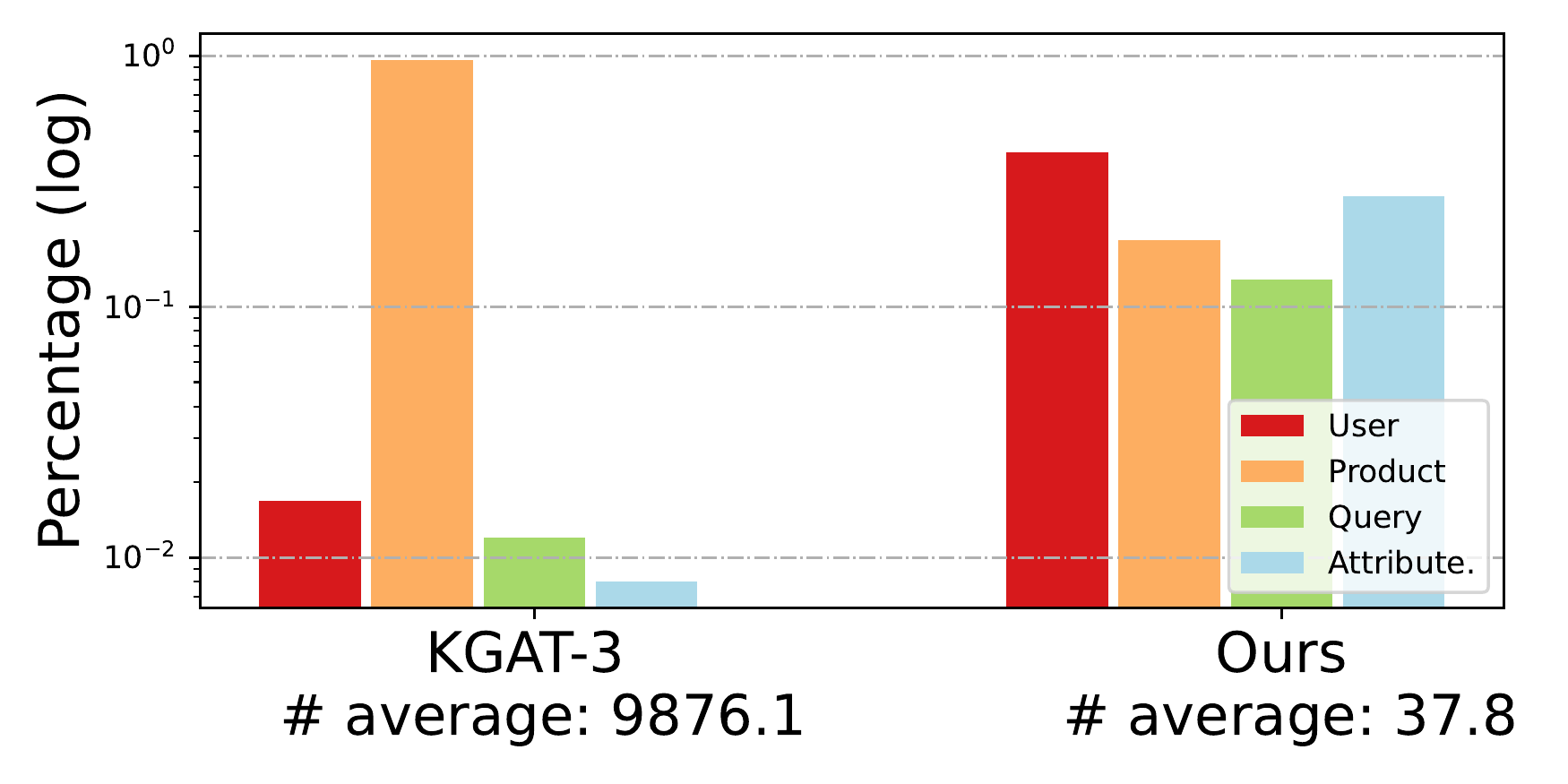}
         \caption{Entity type distribution.}
         \label{fig:retrievaldistribution}
    \end{subfigure}
    \caption{Retrieval analysis. RETE manages to improve performance by stacking more layers, and it can retrieve much more balanced information via a reasonable number of retrieved entities.}
    \label{fig:retrieval_details}
\end{figure}

We analyze how our ensemble subgraph sampler can improve retrieval results by collecting related higher-order entities and filtering out a large ratio of noises. As shown in Figure~\ref{fig:layer}, unlike KGAT, RETE manages to improve performance by stacking more layers (by default, we choose $3$. To investigate the quality of the retrieved entities, Figure~\ref{fig:retrievaldistribution} shows the distributions among entities types as well as average amount of the retrieved entities. RETE can integrate diverse and balanced information via retrieving a reasonable amount of entities. In contrast, KGAT with $3$ layers integrates noisy information from over $9000$ entities for each user, where over $96.4\%$ integrated entities are products.

\subsection{Temporal Analysis and Case Study (RQ4)}
\begin{figure}[t]
    \centering
    \includegraphics[width = 0.95\linewidth]{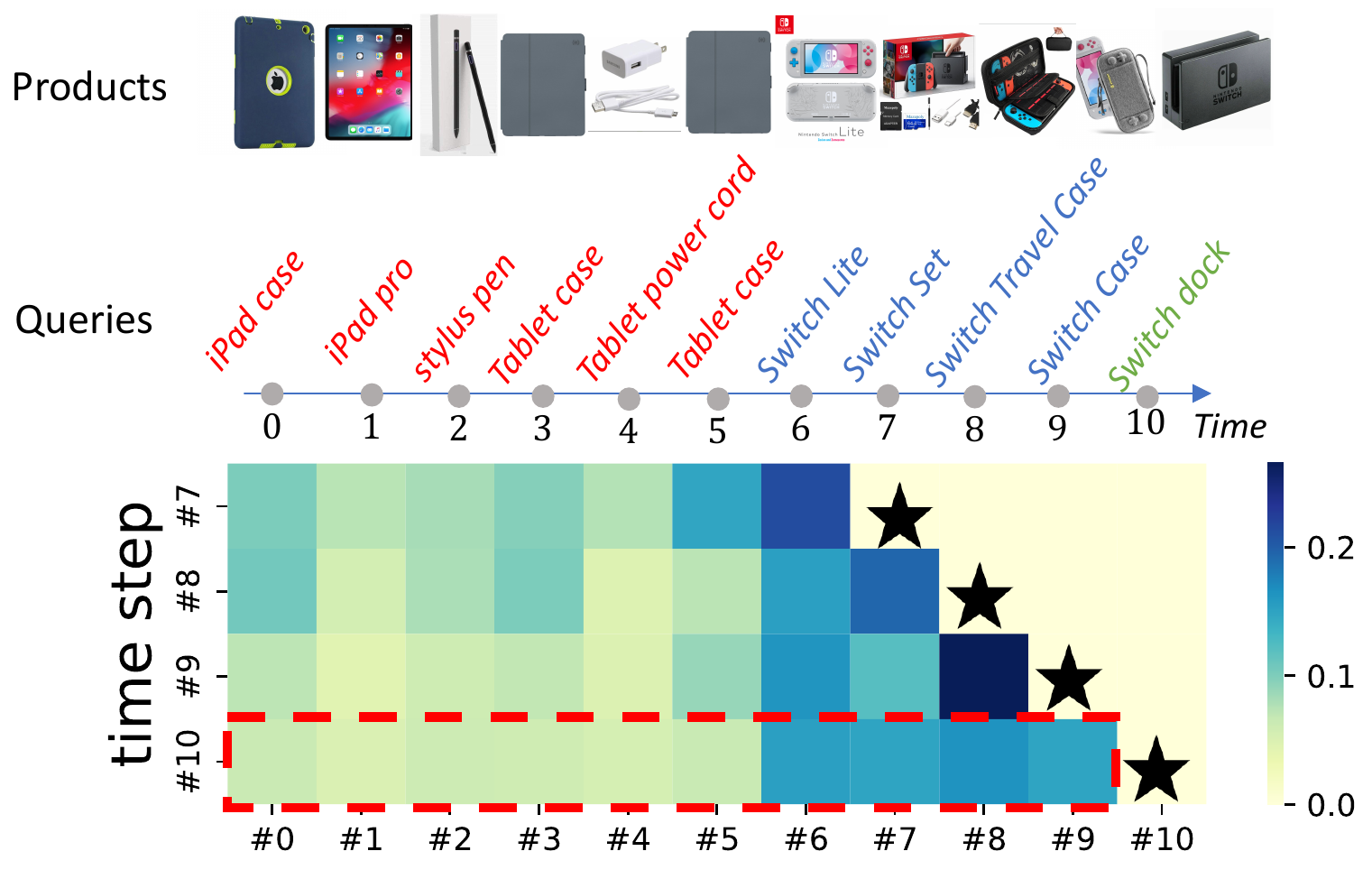}
   \caption{Case study of temporal attention. To forecast the {\color{green}new event} at $t = 10$, it emphasizes more on {\color{blue}related events}, and less on {\color{red}unrelated events}. At each time step, we report and summarize the most frequent event.}
    \label{fig:case}
    \vspace{-3mm}
\end{figure}
To investigate the evolution of user preference and how RETE can capture it, we select one user from the Electronics dataset with $76$ event records. As shown in Figure~\ref{fig:case}, from the most frequent event in each time step, we can observe an obvious interest shift from mobile tablet-related items to Nintendo Switch-related items. Weights before star are used for auto-regressive forecasting. The temporal module can emphasize more on related events and less on unrelated events at new time steps.

\section{conclusion}
\label{sec:conclusion}

In this paper, we explore temporal event forecasting, a new problem considering the temporal influence from both query and product to user behaviors. To enhance the sparse action information of most users and meanwhile capture the evolution of user intents, we propose a novel RETE framework to efficiently retrieve similar entities as subgraphs to enrich the user profile representation and then auto-regressively adapt it to be time-aware. We evaluate the proposed RETE method on both product-centric and query-centric event prediction tasks. Extensive experiments on both public and industrial datasets quantitatively and qualitatively demonstrate the effectiveness of the proposed method. 

\begin{acks}
Research reported in this paper was sponsored in part by DARPA award W911NF-17-C-0099, DARPA award HR001121C0165, Basic Research Office award HQ00342110002, and the Army Research Laboratory under Cooperative Agreement W911NF-17-20196.
\end{acks}

\newpage
\balance
\bibliographystyle{plain}
\bibliography{main.bib}
\clearpage
\appendix
\section{appendix}
\label{sec:appendix}

\subsection{Theoretical analysis}
\label{ap:proof}
Section~\ref{sec:requirement} specifies two model requirements for temporal event forecasting task. In this section, we analysis how RETE satisfies them theoretically. 
\begin{itemize}[leftmargin = 15pt]
    \item \textit{Requirement 1}: It should learn informative and discriminative user representations by considering higher-order information and filtering out large ratio of noise from whole graph.
    \item \textit{Requirement 2}: It should capture user intent evolution from data collected at different time steps, so as to produce up-to-date forecasting.
\end{itemize}

For \textit{Requirement 2}, it is straightforward to show that our temporal attention module can auto-regressively consider new data and update the most up-to-date intents. To analysis \textit{Requirement 1}, we propose the following proposition: 
\begin{proposition}
\label{pro}
For temporal event forecasting task, models able to learn informative and discriminative user representations should satisfy the following necessary conditions: let $m_{[u]}$ denote information gained by user $u$ via $n$-layer of neighbor aggregation, gained information of different users should always be distinguishable: $m_{[u]} \neq m_{[v]}, (u\neq v)$, even under $n \rightarrow \infty$.
\end{proposition}

The correctness of proposition~\ref{pro} is obvious, as models may utilize a large number of layers of neighbor aggregation to integrate higher-order and related information, especially for users with sparse records. So it should be satisfied under $n$ up to $\infty$. To analyze why RETE satisfies the this condition, let $m_{[u]} = \{m_{[u]}^1, m_{[u]}^2, ..., m_{[u]}^t\}$ denote gained information by user $u$ from different time steps. By applying theorem in~\cite{shaDow}, gained information at time step $t$ after $\infty$-layer of GCN-like neighbor aggregation (used in our structural attention module) can be represented as:
\begin{equation}
    m_{[u]}^t = \sqrt{\frac{\delta_{[u]}(u)}{\sum_{v \in \mathcal{G}^t_{[u]}} \delta_{[u]}(v)}} \cdot \boldsymbol{e}_{[u]}^T\boldsymbol{X}_{[u]}
\end{equation}
\noindent
where $\delta_{[u]}(v)$ denotes degree of node $v$ in sampled subgraph $\mathcal{G}_{[u]}^t$, $\boldsymbol{X}_{[u]}$ denotes initial entity embeddings in sampled subgraph $\mathcal{G}_{[u]}^t$, and $\boldsymbol{e}_{[u]}$ denotes eigenvector of $\mathcal{G}_{[u]}^t$ corresponding to largest eigenvalue. By using ensemble subgraph sampler, we can ensure $\mathcal{G}_{[u]}^t \neq \mathcal{G}_{[v]}^t$ even after $\infty$-layer of neighbor aggregation, so that $m_{[u]}^t \neq m_{[v]}^t$ in each time step, and $m_{[u]} \neq m_{[v]}$. Without such subgraph constrains, after large number of neighbor aggregations, the integrated entities easily span the whole graph, i.e., $\mathcal{G}_{[u]}^t = \mathcal{G}_{[v]}^t = \mathcal{G}^t$, making learned user representations much less distinguishable. 
 
\subsection{TransR for product graph learning.}
\label{ap:transr}
Rich meta-data of products forms a heterogeneous product graph $\mathcal{G}_P$, describing important attributes of each product $p \in \mathcal{P}$. Specifically, $\mathcal{G}_P = \{(e, r, e^{\prime})| e \in \mathcal{P}, e^{\prime} \in \mathcal{I} \bigcup \mathcal{Q}, r \in \mathcal{R}_P\}$, where $\mathcal{I}$ denotes attribute set for products, including but not limited to brand, product type and category. $\mathcal{R}_P$ denotes the relation set among them. Each triple $(e, r, e^{\prime}) \in \mathcal{G}_P$ represents a fact indicating that product entity $e$ associate with tail entity $e^\prime$ through relation $r$. $\mathcal{G}_P$ also describe \textit{mapping} relations between products and queries.

To better represent product and query information from static product graph, we utilize a knowledge graph embedding module TransR~\cite{TransR} to project them into the same low-dimensional space. The objective is shown below:
\begin{equation}
    \mathcal{L}_{KGC}=\sum_{(e, r, e^{\prime}) \in \mathcal{G}_{P}} \sum_{\left(e, r, e^{-}\right) \in \mathcal{G}_{P}^{-}} \max \left(0, f_{r}(e, e^{\prime})+\lambda_{KGC}-f_{r}\left(e, e^{-}\right)\right)
\end{equation}
\noindent
where $\mathcal{G}_P$ denotes product graph, $\mathcal{G}_P^-$ denotes negative samples, $\lambda_{KGC}$ is the margin value. Following TransR, we define $f_{r}(h, t)=\left\|\mathbf{h} \mathbf{W}_{r}+\mathbf{r}-\mathbf{t} \mathbf{W}_{r}\right\|_{2}^{2}$, where $\mathbf{W}_{r}$ is trainable relation matrix.

\subsection{Implementation}
\label{ap:implementation}
We implement RETE with Python 3.8.5. We use PyTorch 1.9.1 on CUDA 11.1 to train RETE on GPU. To implement fast subgraph sampling, we adopt methods in~\cite{fastPPR,shaDow} to calculate the approximate PPR value by only traversing the local region around each user. For better efficiency, we implement the sampling part with C++ and the interface between C++ and Python is via PyBind11.

\subsection{Experimental setup}
\label{ap:setup}
\subsubsection{Data collection}

\noindent \textbf{Yelp}. The dataset is adopted in Yelp Challenge 2019 \footnote{https://www.yelp.com/dataset/}, which contains the interaction records between users and businesses like restaurants and bars. We utilize review history between users and businesses as user-product interactions. And we extract discriminative keyphrases from reviews as queries, so as to collect pseudo user-query interactions. We are able to show that jointly consider both pseudo queries and products can improve both performances. We propose the following procedure to collect experimental data.
\begin{enumerate}[leftmargin = 15pt]
    \item We collect data from April 2014 to Jan 2021 and preserve those with ratings higher than $3.0$, as high ratings indicate true user intents. 
    \item To extract keyphrases as pseudo query, we utilize both AutoPhrase~\cite{autophrase}, $2$-gram, and $3$-gram methods to extract keyphrases. Then we calculate TF-IDF scores for each and preserve top $15000$ as pseudo query pools.
    \item We adopt $20$-core setting to collect entities, i.e., we remove users/products/queries with fewer interactions than $20$.
    \item We divide the time span into $28$ time steps according to interaction timestamps. We split them into background/training/val/test (10/10/2/6).
    \item To construct product graph, we first preserve critical attributes like category, location, etc to construct product-to-attribute edge, we collect frequent pairs of businesses from background data as product-to-product edge, we collect query-to-product edge also from background data per each review action.
\end{enumerate}

\noindent\textbf{E-commerce}
We gain access to the search log data including $140$ days and product attribute data. We first collect data under four specific categories: \textit{Electronics}, \textit{Book}, \textit{Music} and \textit{Beauty}. For each category, We propose the following procedure to collect experimental data.
\begin{enumerate}[leftmargin = 15pt]
    \item We collect data from Feb 2021 to June 2021 and preserve those with specific types of actions: click, add cart, follow-on click, and purchase, with ratios of all data $11.8\%$, $5.1\%$, $4.6\%$ and $2.0\%$ respectively. They are preserved as they reveal strong signals of user intents. For each action, we record user, query/product, timestamp, and action type.
    \item We adopt $10$-core setting to collect entities, i.e., we remove users/products/queries with fewer interactions than $10$.
    \item We divide the time span into $28$ time steps according to interaction timestamps. We split them into background/training/val/test (10/10/2/6).
    \item To construct product graph, we first preserve critical attributes like brand, model, etc to construct product-to-attribute edge, we collect frequent pairs of products within the same sessions as product-to-product edge, we collect query-to-product edge also from background data if users have interaction to one product via one search query.
\end{enumerate}

\begin{table}[t]
\caption{Statistics of experimental datasets.}
\label{tb:data}
\footnotesize
\begin{tabular}{c|c|cccc}
\hline
 & \textbf{Public} & \multicolumn{4}{c}{\textbf{Industrial (E-commerce)}} \\ \cline{2-6} 
\textbf{Dataset} & \textbf{Yelp} & \textbf{Electronics} & \textbf{Music} & \textbf{Book} & \textbf{Beauty} \\ \hline
\textbf{\#User} & 22,307 & 5,928 & 7,453 & 37,562 & 47,261 \\
\textbf{\#Product} & 16,153 & 10,129 & 12,105 & 61,215 & 51,686 \\
\textbf{\#Query} & 9,314 & 8,045 & 6,506 & 25,340 & 25,807 \\ \hline
\textbf{\begin{tabular}[c]{@{}c@{}}Product \\ interactions\end{tabular}} & 820,219 & 138,607 & 582,651 & 2,976,112 & 1,385,366 \\
\textbf{\begin{tabular}[c]{@{}c@{}}Query \\ interactions\end{tabular}} & 800,727 & 58,037 & 213,281 & 718,035 & 200,219 \\ \hline
\textbf{\#Entity} & 49,269 & 29,212 & 28,643 & 134,370 & 140,314 \\
\textbf{\#Triplet} & 1,791,788 & 496,701 & 1,832,501 & 7,739,316 & 3,092,010 \\ \hline
\textbf{Time span} & 80 months & 140 days & 140 days & 140 days & 140 days \\
\textbf{\#Time step} & 28 & 28 & 28 & 28 & 28 \\ \hline
\end{tabular}
\end{table}

 We purposefully choose the two platforms because of the various time period. The e-commerce dataset emphasizes more on user short-term interest since the shopping intent is more time-sensitive. By contrast, the Yelp dataset emphasizes more on user long-term interest, since it lasts longer and a user’s choice on businesses is less time-sensitive.To evaluate our framework, we divide the time span into $28$ time steps according to interaction timestamps. We split them into background/training/val/test (10/10/2/6) to train initial entities embeddings (model input), train, validate and test our model respectively. Table~\ref{tb:data} summarizes the statistics of the experimental datasets.

\subsubsection{Baseline}
\begin{itemize}
    \item \textbf{FM}~\cite{FM}. This is a basic factorization model which considers the second-order connections. We construct input features as multi-hot vectors.
    \item \textbf{NFM}~\cite{NFM}. This is a state-of-the-art factorization model, which subsumes FM under neural networks. Specially, we employed one hidden layer to extract features from inputs.
    \item \textbf{GRU4Rec}~\cite{GRU4Rec}. This is a sequential recommendation method that utilizes gated recurrent units (GRU) to learn temporal information of user action sequences.
    \item \textbf{BERT4Rec}~\cite{BERT4Rec}. This is a sequential recommendation method that utilizes BERT to learn temporal information of user action sequences.
    \item \textbf{JODIE}~\cite{JODIE}. This is a dynamic graph method that considers co-evolution of both users and products. We encode rich side information for it via initial embedding.
    \item \textbf{ECFKG}~\cite{ECFKG}. This is an advanced collaborative filtering framework that incorporates knowledge graph for recommendation and utilizes KG embedding to learn representations.
    \item \textbf{KGAT}~\cite{KGAT}. This is an end-to-end knowledge graph attention network that explicitly models the high-order connections and heterogeneous entities and employs an attention mechanism to discriminate the importance of the neighbors.
\end{itemize}

\subsubsection{Setup}
It is worth noting that all baselines are designed for recommending products. To generalize them to be able to predict queries, we train them using product and query loss separately. Those (KGAT, ECFKG) that consider knowledge graph can explicitly fuse information from both product and query. For all methods that need initialization, we utilize a classic and lightweight knowledge embedding method, TransR, to represent multi-relational information in background data, not just IDs. For dynamic methods (BERT4Rec, GRU4Rec, JODIE) that aim to only predict the next interacted product/query, we modify the evaluation protocol to predict all possible products/queries in the following time steps.

For fair comparison, we do not utilize rich semantic information from query entities and product descriptions, as all compared baselines only consider interaction records and structural side information. We fix the dimension of latent vectors of all methods as $128$, and we report the average performance of the best model on the validation set. For RETE, we tune learning rate within $\{0.0001, 0.0005, 0.001,  0.005, 0.01\}$ and regularization weight $\{0.005, \\ 0.05, 0.5\}$ according to $Recall@20$ of product prediction on validation set. We ensemble one PPR sampler and one randomized $3$-hop sampler to retrieve subgraphs, and we stack $3$ layers of graph attentions to better integrate information. 
    
\end{document}